\documentclass[artice]{article}
\usepackage{amssymb,amsmath,amsfonts,amsthm}
\usepackage{setspace}
\usepackage{fullpage}
\usepackage[latin1]{inputenc}
\usepackage{graphicx}
\newtheorem{theorem}{Theorem}
\begin{document}
\title{Improved residuals for linear regression models under heteroskedasticity of unknown form}
\date{}
\author{Andréa V. Rocha$^{a,}$\footnote{E-mail: andrea.rocha@ci.ufpb.br}, Evelina Shamarova$^{b,}$\footnote{E-mail: evelina@mat.ufpb.br}  and Alexandre B. Simas$^{b,}$\footnote{Corresponding author. E-mail: alexandre@mat.ufpb.br}\\\\
\centerline{\small{
$^a$Departamento de Computação Científica, 
Universidade Federal da Para\' iba,
}}\\
\centerline{\small{
$^b$Departamento de Matemática, 
Universidade Federal da Para\' iba,
}}}
\sloppy
\maketitle
\begin{abstract}
In this work we introduce a new residual for normal linear models that are suitable for situations in which we are dealing with heteroskedasticity
of unknown form, they are referred to by principal component analysis (PCA) residuals. 
These residuals are obtained through a linear transformation of the ordinary residuals, by means of a spectral analysis on a
heteroskedasticity-consistent estimator of the covariance matrix. The resulting residuals are independent and normally distributed. 
These residuals provide a simple way to check several assumptions that underlie the normal linear regression model, as well as model adequacy.
Since they are independent and normally distributed, 
one may apply several results on independent random variables directly to these residuals. Finally, we provide an application to real data to illustrate
the usefulness of our residuals.
\end{abstract}
\section{Introduction}
One of the main statistical tools used by practitioners is the regression analysis. From the very definition,
regression models depend on many assumptions, so if one wants to apply a regression model to a particular data set, one must verify if
the assumptions that underlie the statistical model hold. It is well-known that residuals contain important information on such assumptions, 
and therefore play an important role in checking model adequacy. 
The use of residuals
for assessing the adequacy of fitted regression models is nowadays commonplace due to the
widespread availability of statistical software, many of which are capable of displaying residuals
and diagnostic plots, at least for the more commonly used models. 
 Cox and Snell (1968) investigated
residuals in a fairly general class of models and discussed briefly the distribution of these
residuals and the complementary aspect of transforming them, so that they have approximately
the same mean and variance as the ``true'' residuals. Loynes (1969) went further in this direction and provided a transformation so that the resulting residuals
has, approximately, the same distribution as the ``true'' residuals.

One of the most violated assumptions in classical linear regression models is the homoskedasticity assumption. In such case we are dealing with 
heteroskedastic models. The study of normal linear regression models under heteroskedasticity of unknown form is still active. We may cite Cribari-Neto and da Silva (2011),
Cribari-Neto and Lima (2010), Long and Ervin (2000), Flachaire (2005), Godfrey (2006), among others. In this work, we consider the normal linear regression model under heteroskedasticity
of unknown form.

In the classical normal linear regression models, the standard residual, defined as the difference between the response variable and the 
predicted variable is widely used, along with some standardized versions. A drawback found in, essentially, every residual, is that they do 
not form an independent set of random variables. 

Our goal in this work is to introduce a modification of the ordinary residuals in the classical normal linear regression models, under a quite general framework,
namely, under heteroskedasticity of unknown form, in such a way that the 
resulting residuals do form an independent set of random variables. This new set of residuals are suitable for application of several results 
regarding the asymptotics of independent and identically distributed (iid) random variables. Further, they are also suitable to identify 
violation of assumptions in the fitted model. The reason is that one is able to obtain exact results for these residuals. 

To build a set of independent residuals, we apply a classic idea reminiscent of principal component analysis. More precisely, we provide a
new set of residuals that are ranked by their variance, and which are pairwise uncorrelated. The joint normality assumption ensures the independence among them.

The remaining of the article unfolds as follows. In Section 2 we provide a brief review on normal linear models and on heteroskedastic-consistent covariance matrix estimators. We define the new residual 
and deduce its distribution in Section 3. Section 4 contains a discussion on the relevance of the independence among the residuals. In Section 5 
we provide asymptotics on the empirical distribution of the new residuals. Section 6 contains some technical results
regarding estimation of an auxiliary matrix. The usefulness of the residuals is illustrated in Section 7, where we 
present an application to real data. Finally, in Section 8 we provide some concluding remarks.

\section{A brief review on normal liner models and on heteroskedastic-consistent covariance matrix estimators}
Suppose we have a set of $n$ independent random variables $(Y_i,X_{i1},\ldots,X_{ip})$ on $p$ covariates, $i=1,\ldots,n$.
We will occasionally define the covariate $X_{i1} = 1$ to include an intercept. Therefore, we arrive at the model
$$Y_i = \beta_1X_{i1} + \cdots+ \beta_p X_{ip} + \varepsilon_i = \sum_{j=1}^p \beta_j X_{ij} + \varepsilon_i.$$
This identity can be written in vector form as $Y_i = X_i^T \beta + \varepsilon_i$, where $X_i^T = (X_{i1},\ldots,X_{ip})$ and $\beta = (\beta_1,\ldots,\beta_p)$.
In matrix form, this expression becomes
$$Y = X\beta + \varepsilon,$$
where $Y = (Y_1,\ldots,Y_n)^T$ and $\varepsilon = (\varepsilon_1,\ldots,\varepsilon_n)^T$ are $n$-vectors, $X = (X_{ij})$ is an 
$n\times p$ matrix and $\beta$ is the $p$-vector previously defined.

We will assume the following \emph{assumptions} on the model:
\begin{enumerate}
 \item {\bf Linearity}: The regression structure is linear in the parameters;
 \item {\bf Independence and heteroskedasticity of the errors}: $Cov(\varepsilon) = {\rm diag}\{\sigma_1^2,\ldots,\sigma_n^2\} = \Omega$, where $0<\sigma_i^2<\infty$ for $i=1,\ldots,n$;
 \item {\bf Full rank}: The matrix $X$ is of full rank, that is, $rank(X) = p$;
 \item {\bf Normality of the errors}: $\varepsilon \sim N_n(0,\Omega)$, where $N_n(0,\Omega)$ denotes the $n$-multivariate normal distribution
 with mean $0$ and covariance matrix $\Omega$.
\end{enumerate}

Observe that since we assumed $X$ is of full rank, $X^TX$ is invertible. In this case, the ordinary least squares estimator of $\beta$
is given by
$$\hat{\beta} = (X^TX)^{-1}X^Ty,$$
where $y=(y_1,\ldots,y_n)^T$ is a sample of the vector $Y$. 

Note also that, since we are assuming the errors to follow a normal distribution, $\hat{\beta}$ is also a maximum likelihood estimator
and $\hat{\beta} \sim N_p(\beta,(X^TX)^{-1}X^T\Omega X(X^TX)^{-1})$. Let $P = (X^TX)^{-1}X^T$, thus we may write $\hat{\beta} \sim N_p(\beta,P\Omega P^T)$

Observe that, under homoskedasticity, we would have $Cov(\varepsilon) = \sigma^2 I_n$, for some $\sigma^2>0$, and $I_n$ being the $n\times n$ identity matrix.

The predict values are given by
$$\hat{y} = X\hat{\beta} = X(X^TX)^{-1}X^Ty = Hy,$$
where $H = X(X^TX)^{-1}X^T$ is called the \emph{hat} matrix, and we denote its $i$th diagonal element by $h_i$. The hat matrix is also related to the ordinary residuals:
$$\widehat{\varepsilon} \equiv y-\hat{y} = y-Hy = (I_n-H)y.$$
Observe that both $H$ and $I_n-H$ are orthogonal projections. We will now obtain the range and kernel of the hat matrix.
First of all
$$rank(H) = rank(X(X^TX)^{-1}X^T) = rank((X^TX)^{-1}) = rank(X^TX) = p.$$
This implies, in particular, that the column space of the hat matrix coincides with the column space of $X$. Thus,
$$col(H) = col(X) = {\rm span}(X_1,\ldots,X_p),$$
where $X_1 = (X_{11},\ldots,X_{n1}),\ldots,X_p =(X_{1p},\ldots,X_{np})$ are viewed as vectors in $\mathbb{R}^n$, and ${\rm span}(X_1,\ldots,X_p)$ stands for the linear space generated by the $n$-dimensional vectors $X_1,\ldots,X_p$.
Therefore, the orthogonal complement of $col(H)$ is the kernel of $H$. Thus,
$$Ker(H) = \{x\in\mathbb{R}^n; \langle x,X_j\rangle = 0,\quad j=1,\ldots p\},$$
and, in particular, $dim Ker(H) = n-p$.

We we will now provide a review on the heteroskedastic-consistent covariance matrix estimators. 
They are consistent estimators for $Cov(\widehat{\beta})$ under the assumptions made above (including heteroskedasticity of unknown form). They are usually called HC$i$-estimators, where $i$ denotes a version of the 
matrix, the main estimators being: HC0 (White, 1980), HC1 (Hinkley, 1977), HC2 (Horn et al., 1975), HC3 (Davidson and MacKinnon, 1993) and HC4 (Cribari-Neto, 2004).

All of the HC$i$ estimators are given in the form:
$${\rm HC}i =P \widehat{\Omega}_i P^T = P Ewe obtain the spectral decomposition of _i \widehat{\Omega} P^T,$$
where $P = (X^TX)^{-1}X^T$, $\widehat{\Omega} = {\rm diag}\{\widehat{\varepsilon}_1^2,\ldots,\widehat{\varepsilon}_n^2\}$, with $\widehat{\varepsilon}_i$ being the $i$th component
of the ordinary residual vector $\widehat{\varepsilon}$, and the matrices $E_i$, $i=0,1,2,3$ and $4$ are given by
$$E_0 = I_n,\qquad E_1 = \dfrac{n}{n-p} I_n, \qquad E_2 = {\rm diag}\{1/(1-h_i)\}, \qquad E_3 = {\rm diag}\{1/(1-h_i)^2\},$$
and 
$$E_4 = {\rm diag}\{1/(1-h_i)^{\delta_i}\},$$
with $\delta_i = \min\{4,n h_i/p\},$ $i=1,\ldots,n$.

\section{The new residual and its distribution}
In this section we will define our new residuals, which we will call PCA residuals, PCA standing for \emph{principal component analysis}. 
First of all, observe, from the preliminary results, that the residuals satisfy
$$\widehat{\varepsilon} = (I_n-H)y \sim N_n(0,(I_n-H)\Omega).$$

In particular, since $\Omega$ is of full rank, $\widehat{\varepsilon}$ follows a singular multivariate normal distribution. If we let 
$\widehat{\varepsilon}=(\widehat{\varepsilon}_1,\ldots,\widehat{\varepsilon}_n)^T$, it is clear that 
$\widehat{\varepsilon}_i$ and $\widehat{\varepsilon}_j$ are not independent if $i\neq j$. Our goal is to define a new set of residuals that are independent. 

First of all, let us study the spectral decomposition of $I_n-H$. This corresponds to the homoskedastic scenario. Suppose $v$ is an eigenvector of $I_n-H$ associated to the
eigenvalue $1$. Thus
$$(I_n-H)v = v \Longleftrightarrow Hv = 0.$$
Thus, $v$ is an eigenvector associated to the eigenvalue 1 if and only if $v$ is in the kernel of $H$. Since $dim Ker(H) = n-p$,
we have that the eigenspace associated to eigenvalue 1 has dimension $n-p$. 

This gives us a hint of how things should work. Now, we move to the heteroskedastic scenario. By noticing that $\Omega$ has full rank, and since we know $rank(I_n-H)=n-p$,
it follows that $rank[(I_n-H)\Omega] = n-p$. Thus, $0$ is an eigenvalue of the covariance matrix
$$Cov(\varepsilon) = (I_n-H)\Omega,$$
and the dimension of the eigenspace associated to 0 equals $dim Ker(I_n-H) = p$. Unfortunately, under the heteroskedasticity assumption the remaining eigenvalues might be pairwise
different (unlike the homoskedastic case, in which the remaining eigenvalues are all equal to $\sigma^2$). Therefore, we have
the following spectral decomposition of $(I_n-H)\Omega$:
$$(I_n-H)\Omega = Q\Lambda Q^{-1},$$
where $\Lambda = {\rm diag}\{\overbrace{\lambda_1,\ldots,\lambda_{n-p}}^{n-p},\underbrace{0,\ldots,0}_{p})$, and $Q$ is an orthogonal matrix of eigenvectors of $(I_n-H)\Omega$.

Thus, let us define the PCA residuals as
\begin{equation}\label{pcares}
R = Q\widehat{\varepsilon}, 
\end{equation}
where $Q$ is an orthogonal matrix of eigenvectors of $(I_n-H)\Omega$ as defined above. The PCA residuals enjoy the following property
$$R \sim N_n(0,\Lambda),$$
so if we let $R=(R_1,\ldots,R_n)^T$, the residuals $R_i$ and $R_j$ are independent if $i\neq j$. Furthermore,
$R_n,R_{n-1},\ldots,R_{n-p+1}$ are equal to $0$, since they are random variables having zero mean and zero variance, whereas
$R_1,\ldots,R_{n-p}$ are independent random variables following a $N(0,\lambda_i)$ distribution.

We call these residuals PCA, because the idea of diagonalizing the covariance matrix and using its eigenvectors to rank the variance 
are reminiscent of principal component analysis.

Therefore, a good, well-specified model, should be such that the residuals will rank the variances accordingly, i.e., by construction the last $p$ 
residuals will be zero, whereas the remaining should follow independent mean-zero normal distributions. There are several diagnostic plots 
to check if the residuals $R_1,\ldots,R_{n-p}$ are independent and normally distributed random variables. 

In the homoskedastic case, the residuals $R_1,\ldots,R_{n-p}$ will be independent, each following $N(0,\sigma^2$ distribution. The remaining will be equal to zero. 
Therefore, the residuals $R_1,\ldots,R_{n-p}$ will be independent and identically distributed. In such case, one could easily use any of the standard normality tests, 
such as Lilliefors (which is based on the Kolmogorov-Smirnov statistic), Anderson-Darling, Cramer-von-Mises, etc.
Another naive option is to plot a quantile-quantile plot (Q-Q Plot) of these residuals against the quantiles of a 
$N(0,\widehat{\sigma}^2)$ distribution, where $\widehat{\sigma}^2$ is some consistent estimator of $\sigma^2$. 

One should notice that these quantile-quantile plots should be more accurate than those in ordinary residuals, since these residuals are
independent.

In the homoskedastic case, since $\sigma^2$ is unknown, one should consider a strongly consistent estimator. It is immediate, from the strong law of large numbers, that
$$\widehat{\sigma}^2 = \frac{1}{n-p}\sum_{i=1}^{n-p} R_i^2,$$
is strongly consistent. Indeed, one can easily check that this is, in fact, the standard bias-corrected Pearson-based variance estimator, that is
$$\widehat{\sigma}^2 = \frac{1}{n-p}\sum_{i=1}^{n-p} R_i^2 = \frac{1}{n-p}\sum_{i=1}^n (y_i-\widehat{y}_i)^2.$$

We will now define a standardized version of the PCA residual for the homoskedastic case. First, let
$$\widehat{\sigma}_i^2 = \frac{1}{n-p-1}\sum_{\substack{j=1\\ j\neq i}}^{n-p} R_j^2,$$
and observe that $R_i$ and $\widehat{\sigma}_i^2$ are independent.

One should observe that for each $i=1,\ldots,n-p$, $(n-p-1)\widehat{\sigma}_i^2/\sigma^2\sim \chi_{n-p-1}^2$, whereas $R_i/\sigma\sim N(0,1)$, and thus, the standardized version of the PCA residual
$$R_i^\ast = \frac{R_i}{\widehat{\sigma}_i} = \frac{R_i/\sigma}{\widehat{\sigma}_i/\sigma} \sim t_{n-p-1},$$
where $\widehat{\sigma}_i = \sqrt{\widehat{\sigma}_i^2}$, and $t_{n-p-1}$ stands for a Student's-$t$ distribution with $n-p-1$ degrees of freedom.

Therefore, instead of building Q-Q Plot of the PCA residuals against the theoretical quantiles of a $N(0,\widehat{\sigma}^2)$ distribution, 
an \emph{exact} strategy is to build Q-Q Plots of the standardized PCA residuals, namely $R_i^\ast$, against the theoretical quantiles of a $t_{n-p-1}$ distribution.

In a similar fashion, we may define a standardized version of the PCA residual for the heteroskedastic case. In this case,
observe that if
$$(I_n-H)\Omega = Q\Lambda Q^{-1},$$
then, for $i=1,\ldots,n-p$, we have that the residuals $R_i$ are independent and also $R_i \sim N(0,\lambda_i)$. Therefore,
we may define its standardized version as
$$R_i^\ast = \dfrac{R_i}{\sqrt{\lambda_i}},\quad i=1,\ldots,n-p.$$

Since $\lambda_i$ is unknown, it will have to be estimated. Estimation of $\lambda_i$, and more generally of $\Omega$,
is discussed in Section \ref{estimation}.

\section{On the independence among the residuals}
The problem of obtaining independent residuals has already been addressed in the literature with little success. For instance, Loynes (1969) introduced a modified residual having, to order $o(n^{-1})$ the same distribution as the true residuals (that is, the same distribution as the error who motivated the definition of residual, as per Cox and Snell's (1968) seminal paper). However, Loynes argued in the same paper that there was nothing that could be done with respect to the dependence among the residuals. The problem in his approach is the following: he was defining a modification of the residual $R_i$ by considering $R_i' = R_i + \rho_i(R_i)$, where the function $\rho_i(\cdot)$ does not take into consideration the remaining residuals. Thus, on general grounds, it is clear that it is not possible to rule out the correlation among the residuals without taking all the residuals them into consideration. 

Loynes (1969, p. 104) presented the following example to show the difficulty in obtaining independent residuals. Let $\epsilon_i\sim N(0,\sigma^2)$ and $Y_i = \beta + \epsilon_i$. Thus, we find $e = Y_i-\overline{Y} = \epsilon_i - \overline{\epsilon}$, where the over bar indicates that the quantity has been averaged. In this case, it is easy to see that $H = 1/n \boldsymbol{1}_{n\times n}$, where $\boldsymbol{1}_{n\times n}$ is the $n\times n$ matrix with all entries given by 1.

Therefore, $I_n-H$ has rank $n-1$, and by a simple calculation one sees that its kernel, which is the eigenspace associated to the eigenvalue 0, is given by the space $\mathbb{K} = \{t(1,\ldots,1); t\in\mathbb{R}\}$. The remaining eigenvectors, which are associated to the eigenvalue 1, can be obtained by any orthonormal basis 
of $\mathbb{K}^\perp$.

In terms of our PCA residual, this means that the residual equal to zero is given by $(1/\sqrt{n},\ldots,1/\sqrt{n})e = 1/\sqrt{n}\sum_{i=1}^n e_i$. One can find that it is the case by explicit computation. 

The remaining residuals are formed by any orthonormal basis orthogonal to the vector $(1,\ldots,1)$. They are independent and identically normally distributed with mean 0 and variance $\sigma^2$. 

Thus, to obtain the non-zero independent residuals, one must ``exclude'' the principal component given by $(1/\sqrt{n},\ldots,1/\sqrt{n})$, and consider the remaining principal components. 

One must notice that Loynes suggested a linear procedure from a very simple symmetric linear transformation, namely, to consider $e_i + ae_i + b\sum_{i\neq j} e_j$, and showed that it did not work. Our approach goes along the same line, but we considered a much more complex linear transformation.

An alternative manner to obtain independent residuals, with some serious drawbacks, is to consider an equivalent approximation of the BLUS residuals of Theil (1965, 1968). More precisely, consider the estimator $\widetilde{\beta} = \widehat{\beta} + \delta$, where $\delta\sim N(0,\sigma^2(X^\top X)^{-1})$, with $\delta$ being independent of $Y$. Then, by defining the residuals $\widetilde{e} = Y - X\widetilde{\beta}$, one obtains $\widetilde{e} \sim N(0,\sigma^2 I_n)$, thus $\widetilde{e}_1,\ldots,\widetilde{e}_n$ are independent. The main, and most problematic, drawback of this procedure is that $\sigma^2$ must be known.

A drawback on our approach with respect to the approach given by Loynes (1969) is that our residuals are suitable to check model adequacy, normality of the residuals, misspecification of the regression structure, etc., but is \emph{does not} work to identify outlying observations, since the resulting residuals are given by a mixture of the original residuals, thus losing its correspondence with the original observation.
\section{Empirical distribution of the PCA residuals}

A consequence of our approach is that it allows one to easily obtain the asymptotic behavior of the empirical distribution by applying the classical results by Durbin (1973). More precisely, Durbin studied the asymptotic behavior of empirical distribution functions of random samples, under a given sequence of alternative hypotheses that may include the null hypothesis as particular case, with estimated nuisance parameters (in our case, the nuisance parameter is the variance $\sigma^2$). 

The study of the asymptotic behavior of empirical distribution of the residuals is a key result in order to check the assumption of normality of the errors. The asymptotic behavior of empirical distribution of the ordinary residuals for simple normal regression models was obtained by Mukantseva (1977), where it is shown that it is the same as the asymptotic behavior of an iid normally distributed random sample with mean zero. A fairly general result on asymptotic behavior of residuals was given by Loynes (1980). It is clear that one could easily apply Loynes' (1980) result to obtain the same result of Mukantseva (1977) for multiple normal regression. It should be noted, however, that both results by Mukantseva (1977) and Loynes (1980) could not be obtained from the previous results by Kac et al. (1955) or Durbin (1973), due to the presence of dependence among the residuals, and thus required some non-trivial effort to be proved.

In our setup, we are able to obtain an essentially equivalent result to that of Mukantseva (1977) in a multiple regression context in a fairly trivial manner, in the sense that we may apply Durbin's (1973) result directly.

We begin by introducing some notation. Let $\widehat{\sigma}^2 = \dfrac{1}{n-p} \displaystyle\sum_{i=1}^{n-p} R_i^2$, where $R_i$ is the $i$th PCA residual. Therefore,
$$(n-p)^{1/2} (\widehat{\sigma}^2 - \sigma^2) = (n-p)^{1/2} \frac{1}{n-p} \sum_{i=1}^{n-p}(R_i^2 - \sigma^2) = \frac{1}{\sqrt{n-p}} \sum_{i=1}^{n-p}(R_i^2 - \sigma^2).$$
Thus, the technical assumption (A1) of Durbin (1973) holds with $l(R_i,\theta) = R_i^2-\sigma^2$, with $\theta = (\beta,\sigma^2)$. It is also clear that $L = E[l(R_i,\theta)^2] = 2\sigma^4$. Let $\phi_\sigma(x) = 1/\sqrt{2\pi\sigma^2} \exp(-t^2/(2\sigma^2))$ be the density of a $N(0,\sigma^2)$ variable, and 
$\Phi_\sigma(\cdot)$ be the cumulative distribution function of a $N(0,\sigma^2)$ variable. Thus, in Theorem 1 (Durbin, 1973), $g_2(t) = -\Phi_\sigma^{-1}(t)/\sigma \phi_\sigma(\Phi_\sigma^{-1}(t))$, and $h(t) = \sigma^2 \Phi_\sigma^{-1}(t) \phi_\sigma(\Phi_\sigma^{-1}(t))$.

The following theorem is an immediate Corollary to Theorem 1 in Durbin (1973), considering the null hypothesis, that is $\gamma=0$, in Theorem 1 (Durbin, 1973). 

\begin{theorem}
Let $\widehat{F}_n(x) = \dfrac{1}{n-p} \displaystyle\sum_{i=1}^{n-p} 1\{R_i\leq x\}$ be the empirical distribution function of the non-zero PCA residuals $R_1,\ldots,R_{n-p}$. Then, under the hypothesis that $Y = X\beta + \varepsilon$, with $\varepsilon \sim N_n(0,\sigma^2I)$, $\widehat{F}_n$ converges weakly in $D[0,1]$, as $n\to\infty$, with respect to the Skorohod topology, to a gaussian process $Z(t)$ with zero expectation and covariance function
\begin{eqnarray*}
C(Z(t_1),Z(t_2)) &=& \min(t_1,t_2) - t_1t_2 -2\sigma\Phi_\sigma^{-1}(t_1)\Phi_\sigma^{-1}(t_2)\varphi_\sigma(\Phi_\sigma^{-1}(t_1))\varphi_\sigma(\Phi_\sigma^{-1}(t_2))\\
&+&2\sigma^2\Phi_\sigma^{-1}(t_1)\Phi_\sigma^{-1}(t_2) \varphi_\sigma(\Phi_\sigma^{-1}(t_1)) \varphi_\sigma(\Phi_\sigma^{-1}(t_2)).
\end{eqnarray*}
\end{theorem}

From this result one is able to construct asymptotically valid tests, such as Kolmogorov-Smirnov tests (see, for instance, Haj\'ek and \u{S}id\'ak, 1967). Therefore, test normality of the residuals.

The main message of the above result is that our PCA residuals allows one to easily obtain asymptotic results of the residuals since they are iid random variables. Thus, one can find more useful theorems such that the PCA residuals can be applied to. For instance, we could also obtain a similar result to Theorem 1 by applying the results by Kac et al. (1955).

Finally, we were able to obtain an alternative method, and much simpler, to check the normality assumption, as well as model adequacy of multiple normal linear regression models, than the one provided by Mukantseva (1977).

\section{Estimating the matrix $\Omega$}\label{estimation}
In the homoskedastic case, there is no need to estimate the matrix $\Omega$ in order to obtain the PCA residuals, since in this case, $\Omega = \sigma^2 I_n$, and thus the eigenvectors of $(I_n-H)$
coincide with those of $(I_n-H)\Omega$. 

Nevertheless, in the heteroskedastic case, we do need to estimate $\Omega$ since the spectral decomposition of the matrix $(I_n-H)\Omega$ 
is different from that of $(I_n-H)$. To this end, we will use the results for the heteroskedastic-consistent matrix estimators. More precisely, we will use their estimators for $\Omega$, namely
$$\widehat{\Omega}_i = E_i\widehat{\Omega},$$
for $i=0,1,2,3$ and $4$, where  $\widehat{\Omega} = {\rm diag}\{\widehat{\varepsilon}_1^2,\ldots,\widehat{\varepsilon}_n^2\}$,
$$E_0 = I_n,\qquad E_1 = \dfrac{n}{n-p} I_n, \qquad E_2 = {\rm diag}\{1/(1-h_i)\}, \qquad E_3 = {\rm diag}\{1/(1-h_i)^2\},$$
and 
$$E_4 = {\rm diag}\{1/(1-h_i)^{\delta_i}\},$$
with $\delta_i = \min\{4,n h_i/p\},$ $i=1,\ldots,n$. These matrix estimators are known to perform well under heteroskedasticity of unknown form, even in the presence
of leverage points. Thus, they are the natural choices for such estimation.

Thus, one should perform the spectral decomposition on one of the matrices $(I_n-H)\widehat{\Omega}_i$, by choosing a value
$i\in\{0,1,2,3,4\}$, say
$$(I_n - H)\widehat{\Omega}_i = Q_i\widehat{\Lambda}_i Q_i^T,$$
and obtain the corresponding PCA residuals:
$$R^{(i)} = Q_i\widehat{\varepsilon}.$$

Finally, to build the standardized version of such residuals, we use the diagonal elements of the estimated matrix $\widehat{\Lambda}_i$.




\section{Application to real data}
As an application of the PCA residuals to a concrete model, we consider the dataset \emph{states.dta} obtained in \texttt{http://anawida.de/teach/SS14/anawida/5.linReg/data/states.dta.txt}. It consists of educational data on the US states and District of Columbia. It has been analyzed as an example in the book by Hamilton (2009, p. 176). 

The response variable is given by the mean Scholastic Amplitude Test (SAT) scores, which we will denote by \emph{csat}, whereas the covariates are given by: \emph{expense}, which consists of the expenditure per pupil on primary and secondary education; \emph{percent}, which consists of the percentage of high-school graduates taking SAT; and \emph{high}, which consists of the percentage of adults with high-school diploma.

\begin{figure*}[htb]
\begin{center}
\includegraphics[scale=0.4]{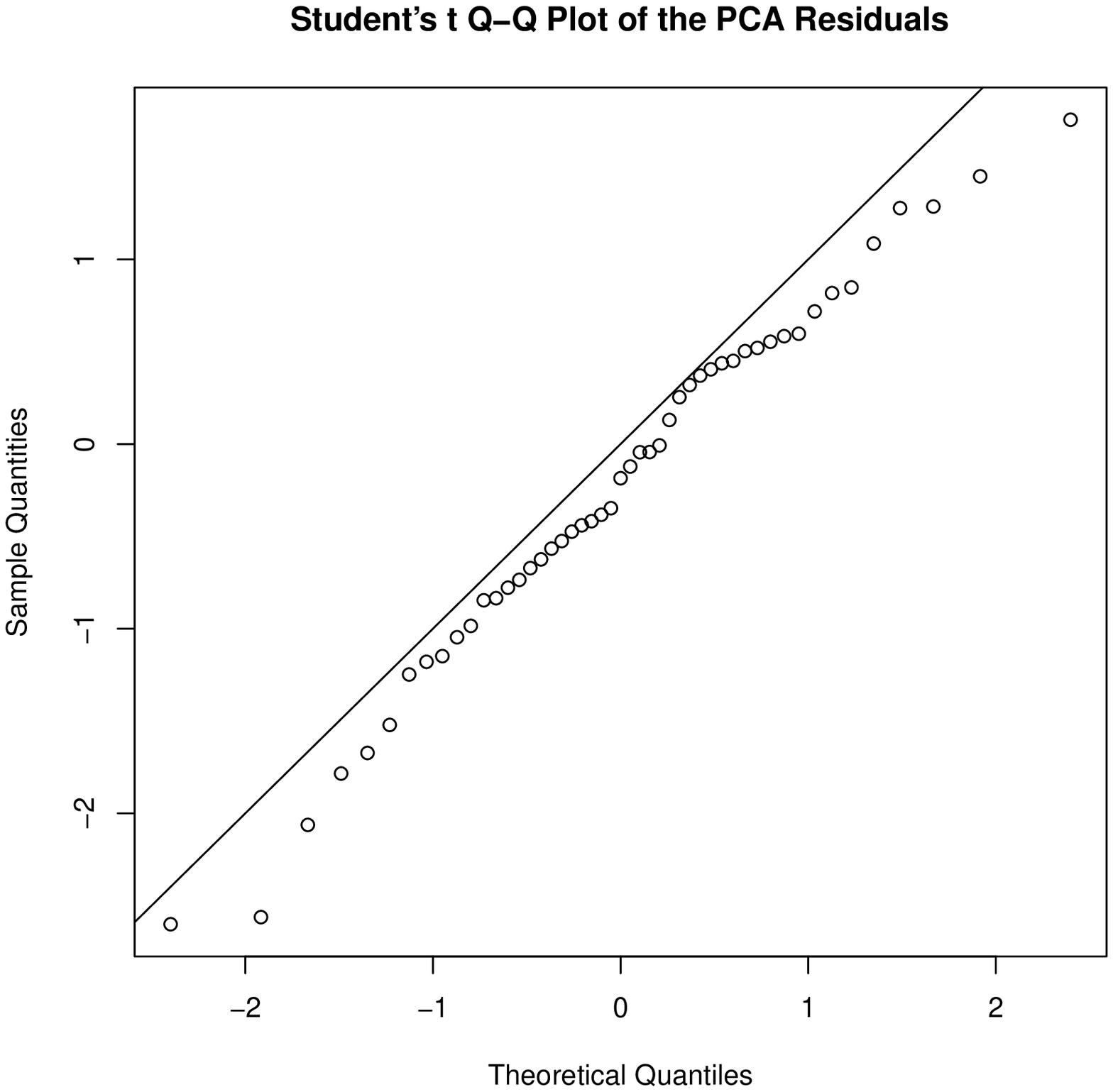}
\includegraphics[scale=0.4]{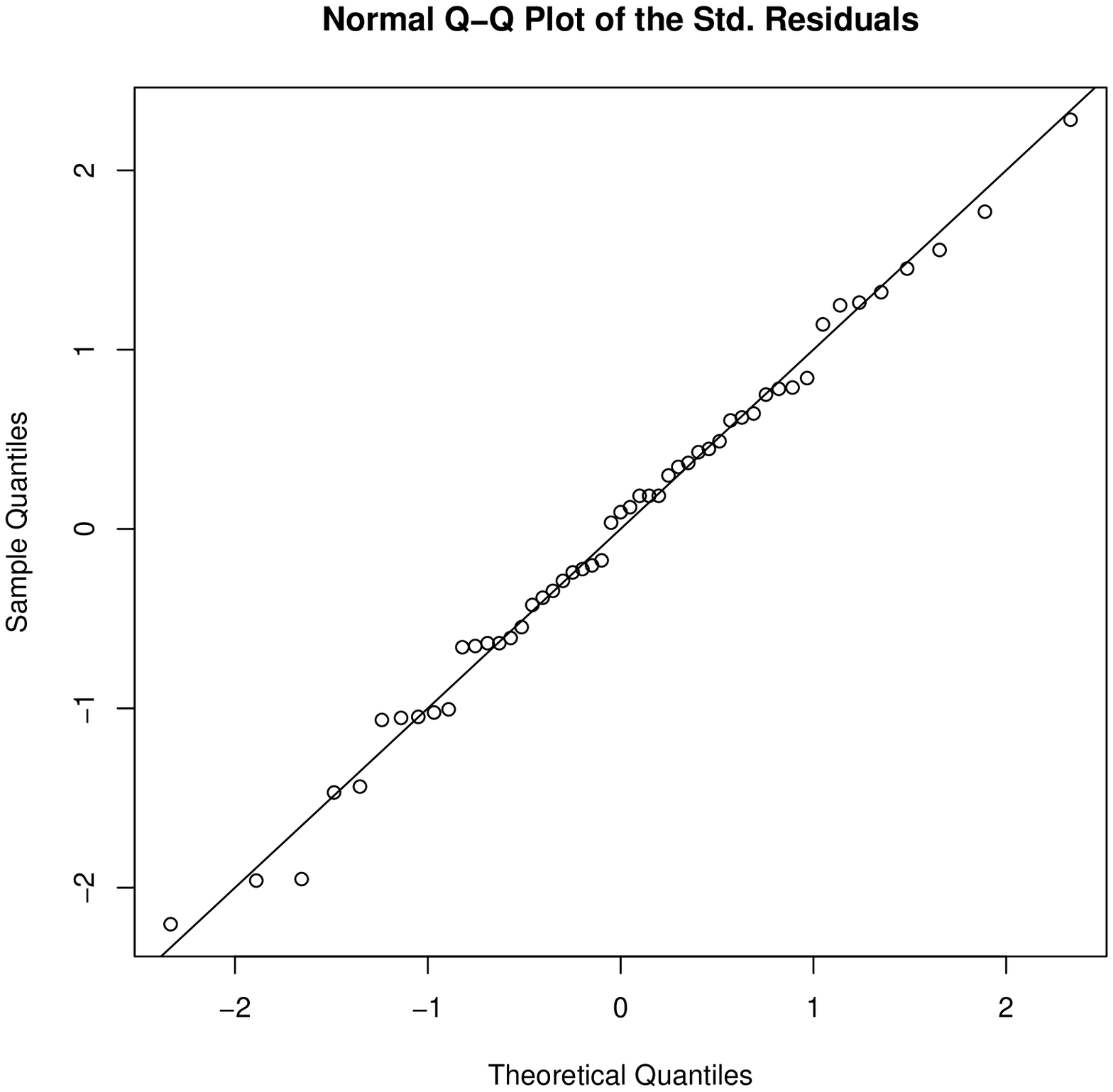}
\caption{Q-Q Plots of the standardized PCA and ordinary residuals for model \eqref{modelharvard}.}
\label{harvard-fig}
\end{center}
\end{figure*}

We shall consider two models to be adjusted. The first one consists of a model suggested in \texttt{http://tutorials.iq.harvard.edu/R/Rstatistics/Rstatistics.html} as an introductory example in \texttt{R}\footnote{See \texttt{http://www.r-project.org/}}, 
where some diagnostic analysis is done, and no indication of violation of the model assumptions is found.
The model is given by
\begin{equation}\label{modelharvard}
csat_i = \beta_0 + \beta_1 expense_i + \varepsilon_i, \qquad i=1,\ldots,51,
\end{equation}
with $\varepsilon_i$ being assumed iid mean-zero normally distributed random variables. 
The sample mean of the \emph{non-zero} PCA residuals for this model is $-13.57$, and its confidence interval 
with $95\%$ significance level is given by
$$\left[ -30.48,3.33\right],$$
which contains zero, thus no violation of the mean-zero assumption of the error is found. 
For this case, we will begin with the PCA residuals under the homoskedasticity assumption. Since, initially,
there is no reason to assume heteroskedasticity. So, we present in Figure \ref{harvard-fig} the standard 
quantile-quantile plots (Q-Q Plots) of 
the standardized PCA residuals against the theoretical quantiles of a Student's $t$ distribution 
with 48 degrees of freedom, and the Q-Q Plots of the standardized ordinary residuals against the theoretical 
quantiles of a standard normal distribution for the residuals in model \eqref{modelharvard}.

The Q-Q Plot of the ordinary residuals suggests an adequate fit. On the other hand, the Q-Q Plot of the PCA residuals 
suggests that the normality assumption is not adequate, since all the points are lying below the diagonal line. 
There is some misspecification on this model. Therefore, our PCA residuals suggests further investigation on this model. 
Going further on this investigation, we applied the augmented Dickey-Fuller test to this model and found a $p$-value 
of $0.0715$, thus not rejecting the null-hypothesis of unit root at the standard confidence level of $95\%$, which 
shows us evidence of non-stationarity on the ordinary residuals. 

Therefore, the PCA residuals, under the homoskedasticity assumption were able to detect departure of stationarity, whereas
the standard residuals were not. Since there was no evidence of heteroskedasticity, we did not consider heteroskedastic PCA residuals
for this model.

\begin{figure*}[hbt]
\begin{center}
\includegraphics[scale=0.4]{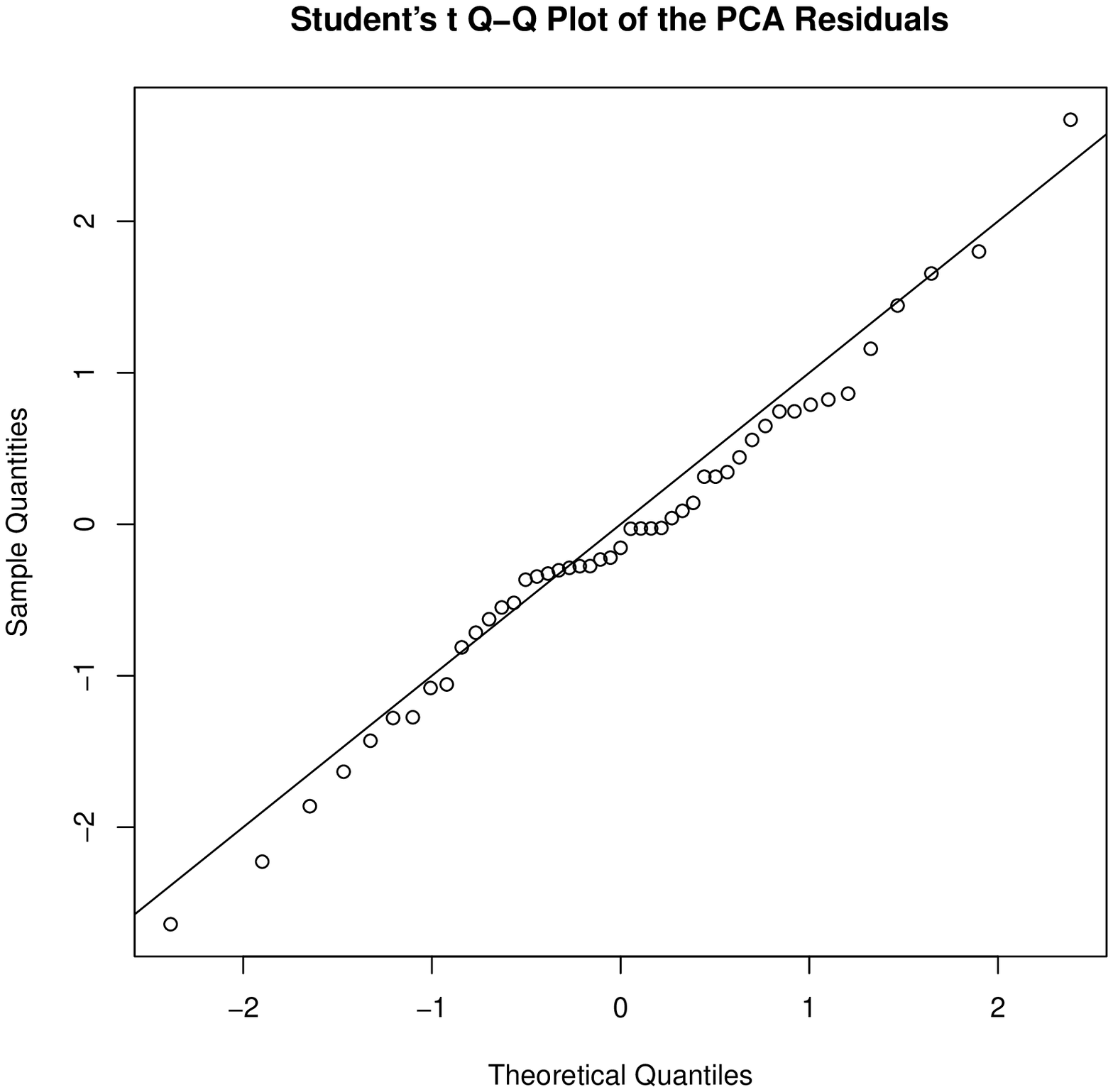}
\includegraphics[scale=0.4]{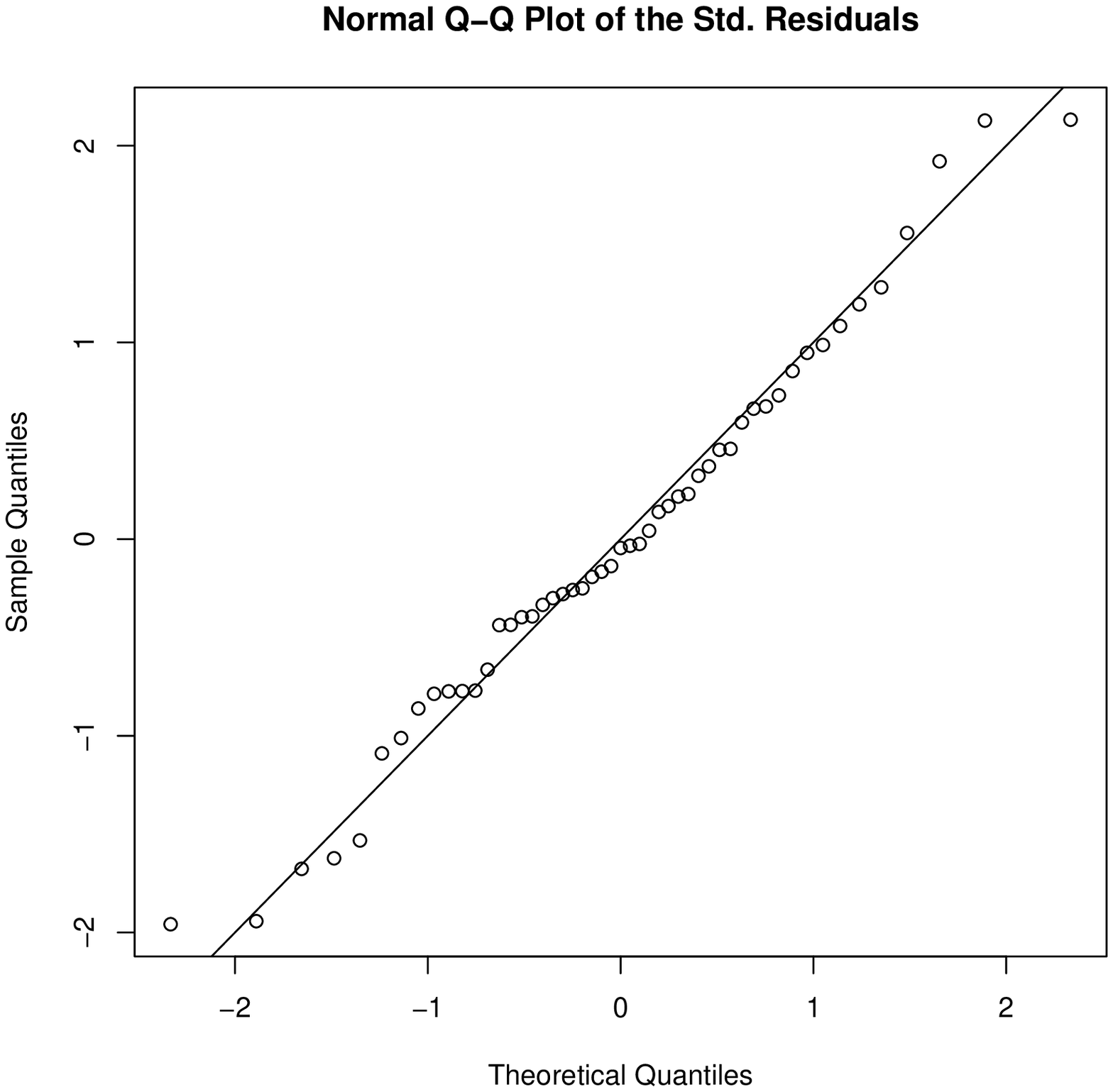}
\caption{Q-Q Plots of the standardized PCA and ordinary residuals for model \eqref{modelprinceton}.}
\label{princeton-fig} 
\end{center}
\end{figure*}

We now consider the model suggested by prof. Torres-Reyna available as an example in a Stata tutorial, namely, 
\texttt{http://www.princeton.edu/$\sim$otorres/Stata/StataTutorial.pdf}. Prof. Torres-Reyna provides a very 
thorough and detailed analysis, also checking various assumptions of the model. The result is that the suggested 
model does violate some of the model assumptions such as homoskedasticity, and also present some colinearity 
between the covariates. So, in particular, this is an heteroskedastic model.
The following model is suggested:
\begin{equation}\label{modelprinceton}
csat_i = \beta_0 + \beta_1 percent_i + \beta_2 high_i + \beta_3 percent_i^2 + \varepsilon_i, \qquad i=1,\ldots,51,
\end{equation}
As for the previous model, we will provide in Figure \ref{princeton-fig} the standard quantile-quantile plots (Q-Q Plots) 
of the standardized PCA residuals built under the homoskedasticity assumption against the theoretical quantiles of a student's 
$t$ distribution with 46 degrees of freedom, together with the Q-Q Plots of the standardized ordinary residuals against the theoretical quantiles of a standard 
normal distribution for the residuals in model \eqref{modelprinceton}. The idea is to observe that the PCA residuals built under 
the assumption of homoskedasticity are able to detect misspecification of the classical model (homoskedastic normal linear 
regression model) whereas the classic residuals are not.

One notices directly that, by looking at Figure \ref{princeton-fig}, and recalling that the Q-Q Plot is exact for this case, we find some, albeit little, evidence of misspecification, or lack of normality (in the sense that even if the residuals follow a normal distribution, they are not identically distributed), but less serious than in the previous model. Thus our PCA residuals agrees with the findings of prof. Torres-Reyna, namely model \eqref{modelprinceton} violates some of the assumptions of the normal linear regression models. 

It is noteworthy that for this model the ordinary residuals suggest, arguably, that the model is adequate, and no violation of the assumptions is found.

\begin{figure*}[hbt]
\begin{center}
\includegraphics[scale=0.6]{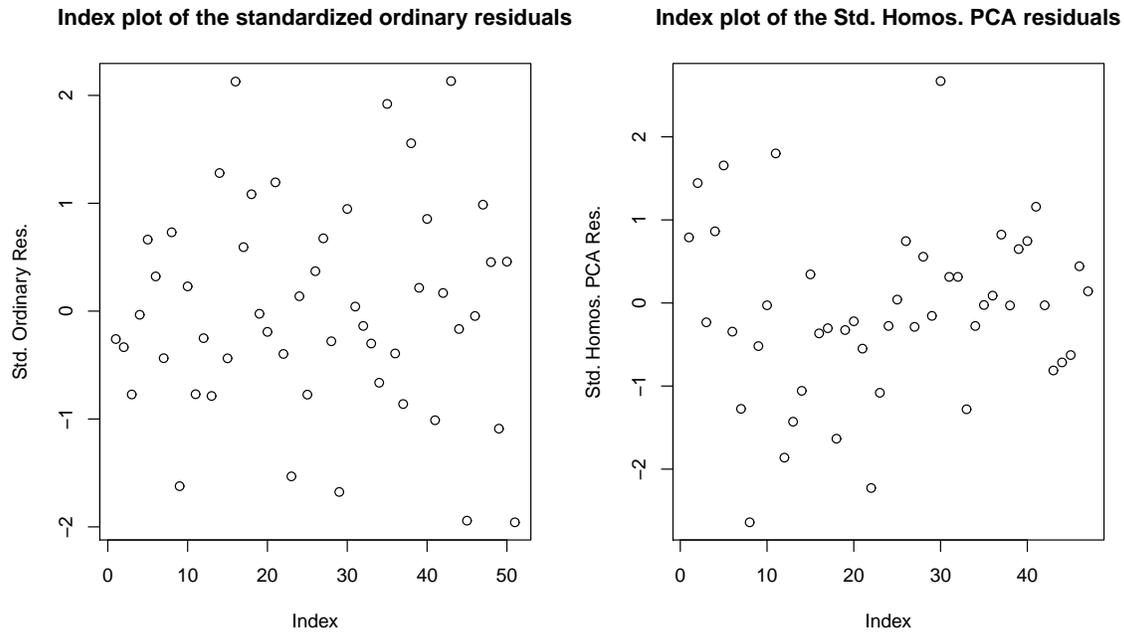}
\caption{Index plots of the standardized ordinary residuals and the standardized homoskedastic PCA residuals for model \eqref{modelprinceton}.}
\label{princeton-fig2} 
\end{center}
\end{figure*}

\begin{figure*}[!hbt]
\begin{center}
\includegraphics[scale=0.6]{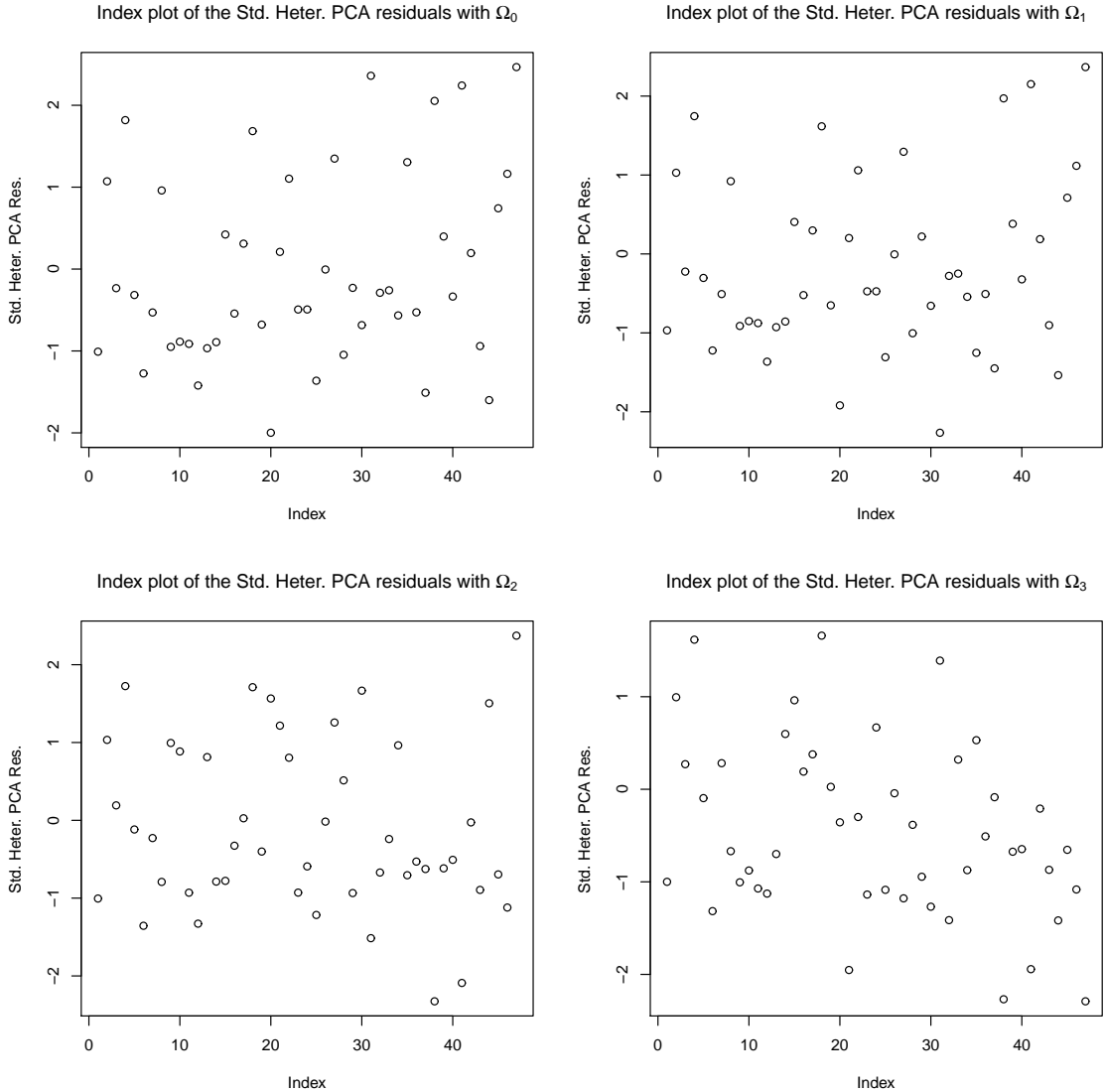}
\caption{Index plots of the standardized heteroskedastic PCA residuals for model \eqref{modelprinceton}.}
\label{princeton-fig3} 
\end{center}
\end{figure*}

Now, we will provide in Figure \ref{princeton-fig2} the index plots of standardized homoskedastic PCA residuals and the standardized ordinary residuals,
whereas in Figure \ref{princeton-fig3} we present the standardized heteroskedastic PCA residuals against their index values.

One should observe that in Figure \ref{princeton-fig2} the pattern of the standardized ordinary residuals
seems uniformly scattered along the rectangle, indicating that there is no misspecification of the model. On the other
hand, the homoskedastic PCA residual has a pattern that indicates the presence of heteroskedasticity, since its variance is somewhat
``varying'' along the indexes. This shows that the homoskedastic PCA residuals are more suitable to identifying violations
on the classical assumptions than the ordinary residuals.

\begin{figure*}[!hbt]
\begin{center}
\includegraphics[scale=0.6]{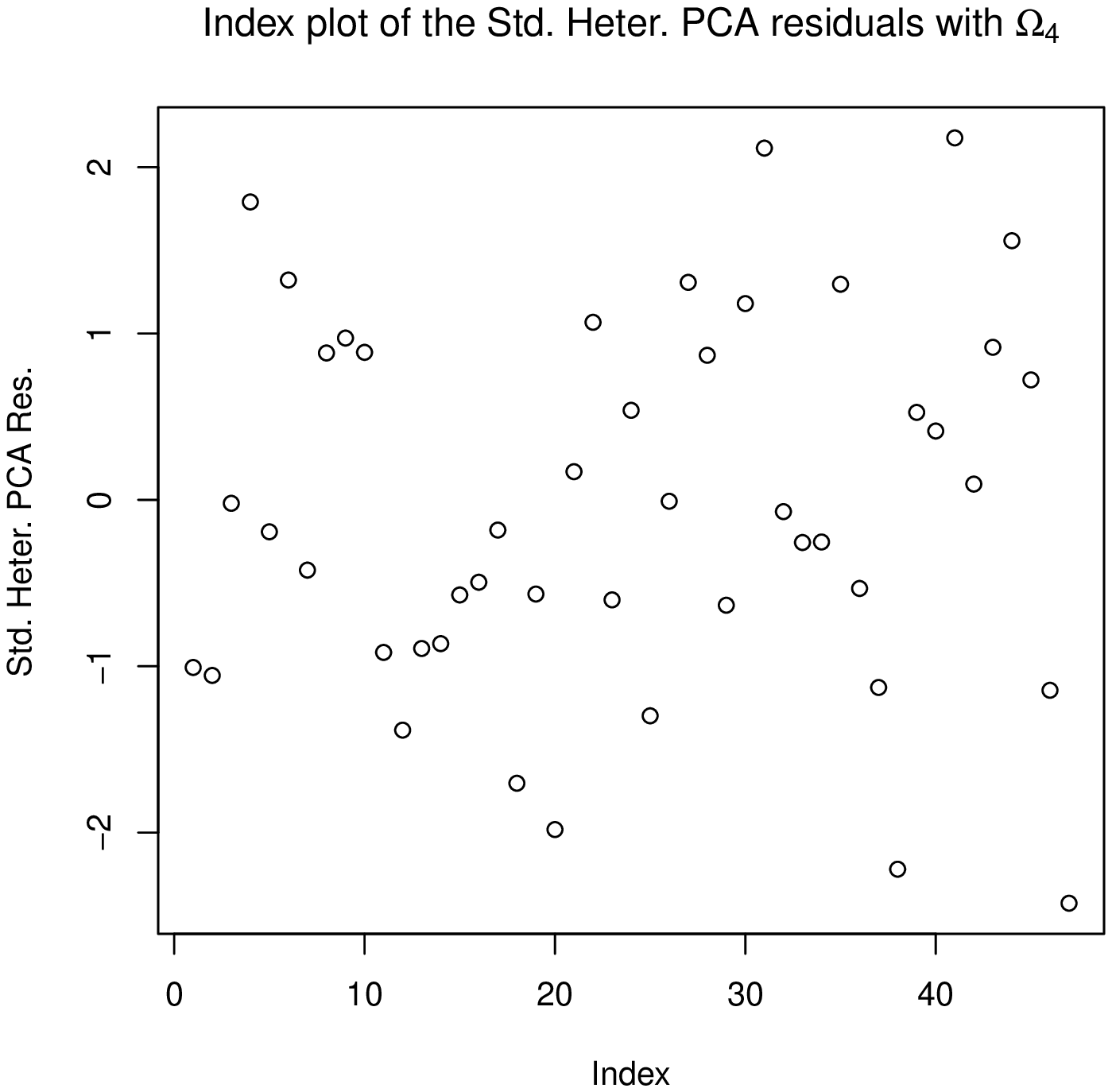}
\caption{Index plots of the standardized heteroskedastic PCA residuals for model \eqref{modelprinceton}.}
\label{princeton-fig4} 
\end{center}
\end{figure*}

We now move to the heteroskedastic PCA residuals to check if there is another violation besides homoskedasticity.
By looking at Figures \ref{princeton-fig3} and \ref{princeton-fig4} we observe a uniform pattern in all versions of the heteroskedastic PCA residuals.
Indicating that, under heteroskedasticity, the model seems well specified, that is, the other assumptions seems to be valid.

This example shows the usefulness of our PCA residuals, since it is a very simple residual, and it was able to identify that both models \eqref{modelharvard} and \eqref{modelprinceton} violated some assumption.
Furthermore, the heteroskedastic version of the PCA residuals for the second model showed that homoskedasticity was apparently the only assumption
violated, which corroborates the findings of prof. Torres-Reyna.



\section{Conclusions}
In this paper we introduced a new residual, which we called the PCA residuals. They have the advantage of being independent, and thus provide an easy way to check the model assumptions. It is noteworthy that our residuals are very simple to obtain and can be easily implemented in the various statistical softwares available.

We showed that the problem of obtaining independent residuals has been addressed in the literature with little success. The main reason is that to obtain independent residuals, we must ``lose'' some residuals. In fact, one must ``lose'' $p$ residuals. But we would like to stress the fact that we \emph{do not} actually lose them, since they are linear combinations of the remaining residuals, and thus their information are still available to us. This shows that it is essentially impossible to modify these residuals and still obtain a set of $n$ independent residuals following a non-singular normal distribution.

Next, we showed that these residuals are suitable to application on theorems for independent and also iid (in the homoskedastic case) random variables, and to illustrate we obtained the asymptotic behavior of its empirical distribution function in a very simple manner. Indeed, by applying a well-known result to our PCA residuals, we were able to essentially find an equivalent result to that of Mukantseva (1977), in the sense that we found a way to test normality in normal linear regression models by studying the asymptotic behavior of empirical distribution of residuals. 


To illustrate the usefulness of our residuals, we presented a dataset previously analyzed, where we found the normal linear regression methodology applied to two different models. We began with a model that was considered adequate, in the sense that no violation of assumptions was found. When we analyzed the PCA residuals for this model, we found evidence of misspecification or lack of normality. Going further on this direction, we found that the residuals had evidence of non-stationarity, thus showing that one or more model assumptions had been violated. We then moved to a different model, that was more thoroughly investigated, and that the person who suggested the model showed that such model suffered from heteroskedasticity and of some colinearity. By analyzing the PCA residuals, we found some evidence of misspecification (less serious than in the previous model), which agreed with the the findings that this model had some model violations. It is interesting to observe that the ordinary residuals seemed adequate for both models, specially the first model. Therefore, these findings show the usefulness of the PCA residuals.

Finally, this strategy is generic enough so that it may be applied to other regression models.
\section*{Acknowledgments} The authors would like to thank CNPq for their financial support.


\begin{thebibliography}{10}
\bibitem{7} Cox, D.R., Snell, E.J. (1968) A general definition of residuals. J. R. Statist. Soc. B. 30, 248-275.
\bibitem{2222111} Cribari-Neto,  F. (2004)  Asymptotic  inference  under  heteroskedasticity  of  unknown  form.  Comput.  Stat.  Data Anal. 45, 215-233.
\bibitem{22331} Cribari-Neto, F., da Silva, W.B. (2011) A new heteroskedasticity-consistent covariance matrix estimator for the linear regression model. AStA Adv Stat Anal 95, 129-146.
\bibitem{2121} Cribari-Neto, F., Lima, M.G.A. (2010) Sequences of bias-adjusted covariance matrix estimators under heteroskedasticity of unknown form. Annals of the Institute of Statistical Mathematics. 62, 1053-1082.
\bibitem{121212} Davidson, R., MacKinnon, J.G. (1993) \emph{Estimation and Inference in Econometrics}. Oxford University Press, New York.
\bibitem{3} Durbin, J. (1973) Weak Convergence of the Sample Distribution Function when Parameters are Estimated. Ann. Stat. 1, 279-290.
\bibitem{33} Flachaire, E. (2005) More efficient tests robust to heteroskedasticity of unknown form. Econometric Reviews. 24, 219-241.
\bibitem{6} Kac, M., Kiefer, J., and Wolfowitz, J. (1955) On Tests of Normality and Other Tests of Goodness of Fit Based on Distance Methods. Ann. Math. Statist. 26, 189-211.
\bibitem{222} Godfrey, L.G. (2006) Tests for regression models with heteroskedasticity of unknown form. Comp. Stat. Data. Anal., 50, 2715-2733.
\bibitem{11} Haj\'ek, J., \u{S}id\'ak, Z. (1967) \emph{Theory of Rank Tests}. Academic Press, New York.
\bibitem{10} Hamilton, L.C. (2008) \emph{Statistics with Stata: Updated for Version 10}. Duxbury Press; 7 edition.
\bibitem{333} Hinkley, D.V. (1977) Jackknifing in unbalanced situations. Technometrics, 19, 285-292.
\bibitem{3311} Horn, S.D., Horn, R.A. (1975) Duncan, D.B.: Estimating heteroskedastic variances in linear models. J. Am. Stat. Assoc., 70, 380-385.
\bibitem{1} Loynes, R.M. (1969) On Cox and Snell's General Definition of Residuals. J. R. Statist. Soc. B. 31, 103-106.
\bibitem{2} Loynes, R.M. (1980) The Empirical Distribution Function of Residuals from Generalised Regression. Ann. Stat. 8, 285-298.
\bibitem{4} Mukantseva, L.A. (1977). Testing normality in one-dimensional and multidimensional linear regression. Theory Probab. Appl. 22, 591-602. 
\bibitem{5} Rocha, A.V., Simas, A.B. (2016) On Pearson residuals in exponential family nonlinear models. Preprint.
\bibitem{8} Theil, H. (1965) The Analysis of Disturbances in Regression Analysis. J. Amer. Statist. Assoc. 60, 1067-1079.
\bibitem{9} Theil, H. (1968) A Simplification of the Blus Procedure for Analyzing Regression Disturbances. J. Amer. Statist. Assoc. 63, 242-251.
\bibitem{2333} White, H. (1980) A heteroskedasticity-consistent covariance matrix estimator and a direct test for heteroskedasticity. Econometrica, 48, 817-838.



\end{thebibliography}
\end{document}